\documentclass[aps,prl,twocolumn,superscriptaddress,showpacs,floatfix]{revtex4}

\usepackage{graphicx,color}
\usepackage{dcolumn}
\usepackage{bm}
\usepackage{stmaryrd}
\usepackage{latexsym}
\usepackage{amssymb}
\usepackage{amsfonts}
\usepackage{amsmath}
\usepackage{xcolor}

\newcommand{\bS}{{\bf S}}
\newcommand{\bD}{{\bf D}}
\newcommand{\bk}{{\bf k}}
\newcommand{\br}{{\bf r}}

\begin{document}

\title{Systematic Search and A New Family of Skyrmion Materials}

\author{Wei Li}
\affiliation{State Key Laboratory of Functional Materials for Informatics and Shanghai Center for Superconductivity, Shanghai Institute of Microsystem and Information Technology, Chinese Academy of Sciences, Shanghai 200050, China}

\author{Jiadong Zang}
\email{jiadongzang@gmail.com}
\affiliation{Institute for Quantum
Matter, Department of Physics and Astronomy, Johns Hopkins
University, Baltimore, Maryland 21218, USA}

\date{\today}


\begin{abstract}
Magnetic skyrmions have recently attracted great attentions. However
they are harbored in very limited numbers of magnets up to now. The
search of new helimagnetic materials is thus an urgent topic in the
field of skyrmion physics. In this letter, we provide a guideline on
this issue, and discuss the possibility of realizing skyrmions in a
new family of molybdenum nitrides $A_2$Mo$_3$N ($A$=Fe, Co, and Rh).
By means of the first-principles calculations, the electronic and
magnetic structures are calculated and the existence of strong
Dzyaloshinskii-Moriya interaction is demonstrated.
\end{abstract}

\maketitle

The magnetic skyrmion is a swirling-like spin texture with
nontrivial topology, where magnetic moments point in all directions
in the space, in contrast to any other trivial textures such as
ferromagnetism. Soon after its first encounter with
magnetism~\cite{rossler_spontaneous_2006}, the skyrmion was observed
in MnSi by neutron scattering~\cite{muhlbauer_skyrmion_2009}, and
later confirmed in Fe$_{1-x}$Co$_x$Si by Lorentz transmission
electron microscopy~\cite{yu_real-space_2010}. It has attracted
great attentions due to its promise of future applications in memory
devices~\cite{fert_skyrmions_2013}. Recent developments have
witnessed skyrmions in several other bulk materials, such as
FeGe~\cite{yu_near_2011}, hosting skyrmions quite close to room
temperatures, and Cu$_2$OSeO$_3$~\cite{seki_observation_2012}, where
skyrmions are insulating and exhibiting multiferroic properties.

However all these materials belong to the same lattice class called
B20 compounds in the Strukturbericht Symbol. Although this class
belongs to the cubic crystal system, complicated distributions of
atoms in each unit cell dramatically bring down the symmetry. Thus
the point group of B20 compounds is the tetartoidal T23 group, the
one with lowest symmetry in the cubic system, where both inversion
and mirror symmetries are missing. It has been well understood that
the inversion symmetry breaking generates the Dzyaloshinskii-Moriya
(DM)
interaction~\cite{dzyaloshinsky_thermodynamic_1958,moriya_anisotropic_1960},
which competes with the Heisenberg exchange, and induces skyrmions
or conical state at finite magnetic fields. This competition is also
the origin of the spin helices under low magnetic
fields~\cite{uchida_real-space_2006}. That is why B20 compounds are
termed as helimagnets. The DM interaction is noticeably large in B20
compounds, so that the dipolar interaction is negligible. This
provides skyrmions therein with the same chirality and controllable
properties. Facing the limited choices of B20 compounds, an urgent
problem thus arises that how to find other helimagnet materials
harboring skyrmions, spin helices, or conical states driven by the
DM interaction.

In B20 helimagnets, the essential Hamiltonian in the continuum limit
is given by
\begin{equation}
H=J(\nabla\bS)^2+D\bS\cdot(\nabla\times\bS)-\mathbf{h}\cdot\bS\label{eq:hamiltonian}
\end{equation}
It has been extensively tested this Hamiltonian well characterizes
the phase diagrams of the B20 compounds. The last term in Eq.
(\ref{eq:hamiltonian}) is the Zeeman coupling. At a finite window of
the magnetic field and temperature, the skyrmion would appear, while
the helix is energetically favored at low magnetic fields. The
presence of the helix can be readily understood by visiting the
leading two terms, the Heisenberg exchange and the DM interaction
respectively, in Eq. (\ref{eq:hamiltonian}). One can perform the
Fourier transformation of the Hamiltonian, and get a quadratic
function in momentum $\bk$ from the Heisenberg exchange, while
receiving a linear function in momentum from the DM interaction. The
Hamiltonian is thus minimized at a finite momentum $\bk_0$, which is
the wavevector of spin helix. It shows a linear term in momentum is
essential.

In details, the Fourier component of the Heisenberg exchange is
given by $H_\text{Hei}(\bk)=J\bk^2|\bS_\bk|^2$, where $\bS_\bk$ is
the Fourier component of the spin $\bS_\bk=\frac{1}{V}\int
d\mathbf{r}\bS(\mathbf{r})\exp(i\bk\cdot\mathbf{r})$. Under
rotations, $\bS$ transforms in the same way as $\bk$. As a result,
$H_\text{Hei}(\bk)$ is rotationally invariant. Furthermore,
inversion and mirror symmetries are also respected. Therefore this
is a generic quadratic term for all ferromagnets.

On the other hand, the DM interaction provides a linear term in
momentum; $H_{\text{DM}}(\bk)=iD\bS_\bk\cdot(\bk\times\bS_{-\bk})$.
Although the rotational symmetry is still preserved, inversion
symmetry is apparently broken, which is the well know precondition
for the DM interactions. However a long overlooked fact is the
mirror symmetry is also broken by this DM interaction. This comes
from the fact that $\bS_\bk\cdot(\bk\times\bS_{-\bk})$ is a
pseudoscalar. Under any improper rotation such as mirror reflection
in the lattice, this term flips sign, and should be ruled out in the
energy. That is why magic is witnessed in B20 compounds. The point
group T23 only has pure rotations, thus allows the DM interaction in
this form. In case a mirror plane exists, the DM interaction arise
from broken inversion symmetry must be staggered, failing to end up
with a continuum limit in this form. Now a question arises whether
this form is the only allowed term linear in $\bk$ for any
materials.

To answer this question, we can rewrite any $\bk$-linear terms as a
tensor product $H_{\text{DM}}(\bk)=id_{ijm}k^iS_{-\bk}^jS_{\bk}^m$,
where $d_{ijm}$ is a third order tensor that can be constructed from
symmetry analysis. Any symmetry operation $R$ can be represented as
a 3$\times$3 matrix in natural basis (x, y, z). Under such
operation, vector $\bk$ transforms as $k_i\rightarrow k_jR_{ji}$,
while the pseudovectors $\bS_{\pm\bk}$ transform as $S_i\rightarrow
|R|S_jR_{ji}$, where $|R|$ is the determinant of $R$ matrix. If $R$
is an improper rotation, $|R|=-1$. Once $R$ is a symmetry operation,
energy should be invariant under such rotation, therefore the tensor
$d_{ijm}$ must satisfy the Neumann's principle:
\begin{equation}
d_{ijm}=R_{ip}R_{jq}R_{mr}d_{pqr}\label{eq:neumann}
\end{equation}
In practice, one does not need to go through all symmetry operations
in order to determine the $d$ tensor. Most operations can be written
as products of some independent matrices, called generating
matrices~\cite{birss_symmetry_1964}, within the same point group.
For T23 point group, the generating matrices are $C_2$ and $C_3$
rotations. The Neumann's principle thus leads to the constrain that
$d_{xyz}=d_{yzx}=d_{zxy}$, and $d_{xzy}=d_{yxz}=d_{zyx}$. One can
symmetrize these parameters by $d_{xyz}=S+D$ and $d_{xzy}=S-D$.
However because the whole Hamiltonian can be reorganized as
$\sum_\bk H_{\text{DM}}(\bk)=\sum_\bk
i(d_{ijm}-d_{imj})k^iS_{-\bk}^jS_{\bk}^m$, the symmetric component
$S$ does not contribute. The resulting Hamiltonian is thus $\sum_\bk
iD\varepsilon_{ijm}k^iS_{-\bk}^jS_{\bk}^m$, which reproduces the DM
interaction in Eq. (\ref{eq:hamiltonian}).

The same method applies to any other lattices. Contribution to the
Hamiltonian from any tensor with redundant indices vanishes when
completing the summation over momenta. The relevant terms are six
components with indices permutations of (x, y, z). For future
convenience, these six components are symmetrized as
$d_{xyz}=\alpha_S+\alpha_A$, $d_{yxz}=\alpha_S-\alpha_A$,
$d_{yzx}=\beta_S+\beta_A$, $d_{xzy}=\beta_S-\beta_A$,
$d_{zxy}=\gamma_S+\gamma_A$, $d_{zyx}=\gamma_S-\gamma_A$. As a
result, the total Hamiltonian is given by
\begin{eqnarray}
H=\int d^3\br&&[J(\nabla\bS)^2-\mathbf{h}\cdot\bS\nonumber\\
+&&\frac{1}{2}(\alpha_S+\alpha_A-\beta_S+\beta_A)\bS\cdot(\partial_{\hat{x}}\times\bS)\nonumber\\
+&&\frac{1}{2}(-\alpha_S+\alpha_A+\beta_S+\beta_A)\bS\cdot(\partial_{\hat{y}}\times\bS)\nonumber\\
+&&\gamma_A\bS\cdot(\partial_{\hat{z}}\times\bS)]
\end{eqnarray}
where $\partial_{\hat{r}}$ is the directional derivative along $\br$
direction. Under low magnetic field $\mathbf{h}$, spin helix is thus
formed along certain directions given the competition between
anisotropic DM interaction and the Heisenberg exchange.

\begin{table}
\caption{Constrains of nonzero $d_{ijk}$ parameters for all possible
point groups}
\begin{tabular}{ccc}
\hline \hline
  Class & Constrains  &   Point Groups   \\
\hline
  I &   No Constrain  & $C_1$, $C_2$, $D_2$   \\
  II    &   $\alpha_{S}=\beta_S=\gamma_S=0$ & $C_4$, $D_4$, $C_3$, $D_3$, $C_6$, $D_6$  \\
  III   &   $\alpha_A=\beta_A=\gamma_A=0$ & $S_4$, $D_{2d}$  \\
  IV    &   $\alpha_{S}=\beta_S=\gamma_S$, $\alpha_A=\beta_A=\gamma_A$    & T     \\
  V &   $\alpha_{S}=\beta_S=\gamma_S=0$, $\alpha_A=\beta_A=\gamma_A$    & O     \\
\hline \hline
\end{tabular}\label{table_pointGroups}
\end{table}

A complete list of point groups contributing to nonzero DM
interactions is summarized in Table \ref{table_pointGroups}, where
the B20 compounds are located in class-IV. We know that the point
group can be decomposed into two categories, the ones with and
without improper rotations. Most groups in Table
\ref{table_pointGroups} belong to the category without any improper
rotations. Mirror, in addition to inversion, is broken in these
lattices. The only exception is class-III; $S_4$ and $D_{2d}$
groups, where $\bS\cdot(\partial_{\hat{z}}\times\bS)$ is prohibited.
Helices or skyrmions can only form in the plane perpendicular within
the horizontal mirror. In contrast, class-II allows the presence of
the DM interaction in the $z$-direction. The DM interaction of this
class can be rewritten as
$\frac{1}{2}(\alpha_A+\beta_A+\gamma_A)\bS\cdot(\nabla\times\bS)+\frac{1}{2}\gamma_A\bS\cdot(\partial_{\hat{z}}\times\bS)$.
It shows explicitly that the $z$-direction is distinct from the $xy$
plane. Although helices or conics can propagate along $z$-direction,
the skyrmion crystal would favor to accommodate in the $xy$ plane
instead. However one should be aware that the spin anisotropies are
quite large in these lattices with reduced symmetry, which are
enemies of the skyrmion. One can also construct spin anisotropies by
the symmetry tensors up to arbitrary order. These details are out of
the scope of this work. In the future, case by case studies can be
conducted for specific materials.

\begin{figure}[tbp]
\includegraphics[width=9cm, height=4cm]{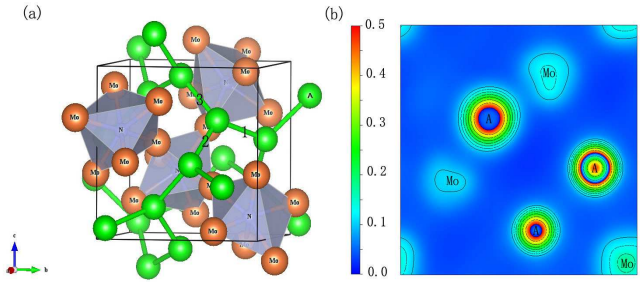}
\caption{(Color online) (a) The schematic crystal structure of the
filled $\beta$-manganese structural $A_2$Mo$_3$N ($A$=Co, Fe, and
Rh). Space within the (10, 3)-a network of $A$ atoms (green) is
filled by vertex-sharing Mo$_6$N. (b) The charge density
distribution in the (001) plane crossing the $A$-Mo-$A$ atoms in the
NM  state calculations.}\label{fig1}
\end{figure}

The most important message delivered from Table
\ref{table_pointGroups} is class-V, which is described by {\it
exactly the same} Hamiltonian as the B20 compounds in class-IV. The
spin anisotropies are also the same in these two classes. Therefore
we expect the spin physics observed in B20 compounds are also
persistent in the point group $O$. There exists 8 space groups in
$O$ group. We found that the most promising material in analogy to
B20 compounds is the $A_2$Mo$_3$N family with $A$=Fe, Co, Rh, or
their alloys. The pure
$A_2$Mo$_3$N~\cite{TJPrior,TJPrior2,PDBattle}, shown in
Fig.~\ref{fig1}(a), has the filled $\beta$-manganese structure,
where $A$ atoms lie on the $8c$ positions of a cubic unit cell with
space group symmetry $P4_132$ forming a single (10, 3)-a
network~\cite{AFWells}. The space within this network is filled by
corner-shared Mo$_6$N octahedra. This structure is entirely
analogous to that of $\beta$-manganese, differing only by the
addition of the interstitial non-metal atoms. We expect large
strength of spin-orbital coupling in this family, and thus skyrmions
can be hosted.

To quantitatively confirm the aforementioned conjecture of promising
material, we carried out the first-principle calculations of pure
$A_2$Mo$_3$N using the projected augmented wave method as
implemented in the VASP code~\cite{VASP}, where the
exchange-correlation potential was calculated using the generalized
gradient approximation (GGA) as proposed by Pedrew, Burke, and
Ernzerhof (PBE)~\cite{PBE}. All atomic positions and lattice
constants of $A_2$Mo$_3$N were allowed to relax simultaneously to
minimize the energy. A 500 eV cutoff in the plane wave expansion
ensures the convergence of calculations up to $10^{-5}$ eV, and all
atomic positions and the lattice constants were optimized until the
largest force on any atom was 0.005 eV/\AA. Furthermore, we used a
$8\times 8\times 8$ Monkhorst-Pack k-grid Brillouin zone sampling
throughout all of calculations. In addition, the spin-orbit coupling
was also included with the second variational method. The lattice
constants and the internal coordinates of the atomic positions for
the systems of $A_2$Mo$_3$N are all optimized and listed in
Table~\ref{table1}, which shows that the optimized lattice
parameters for both Fe$_2$Mo$_3$N and Co$_2$Mo$_3$N are quite
consistent with that from
experiments~\cite{TJPrior,TJPrior2,PDBattle}. Although lattice
parameters for Rh$_2$Mo$_3$N are lacking experimentally, our
optimized values are larger than those of Fe$_2$Mo$_3$N and
Co$_2$Mo$_3$N, which are reasonable as the atomic radii of Rh is
apparently larger than that for Fe and Co.

\begin{table}[b]
\caption{The optimized lattice constants as well as the internal
coordinates for the filled $\beta$-manganese structured $A_2$Mo$_3$N
($A$=Co, Fe, and Rh).}
\begin{tabular}{cccc}
\hline
\hline
~~~~~$A_2$Mo$_3$N~~~~  & ~~~~~$A$=Fe~~~~~ & ~~~~~$A$=Co~~~~~ & ~~~~~$A$=Rh~~~~   \\
\hline
  a (\AA)  & 6.6458 & 6.6367 & 6.8224  \\
\hline
  $A$ ($\hat{x}=\hat{y}=\hat{z}$)  & 0.0723 & 0.0677 & 0.0598  \\
  Mo ($\hat{y}$) & 0.2019 & 0.2002 & 0.2021  \\
  Mo ($\hat{z}$) & 0.4519 & 0.4502 & 0.4521  \\
\hline
\hline
\end{tabular}\label{table1}
\end{table}

\begin{figure}[tbp]
\includegraphics[width=9cm, height=7cm]{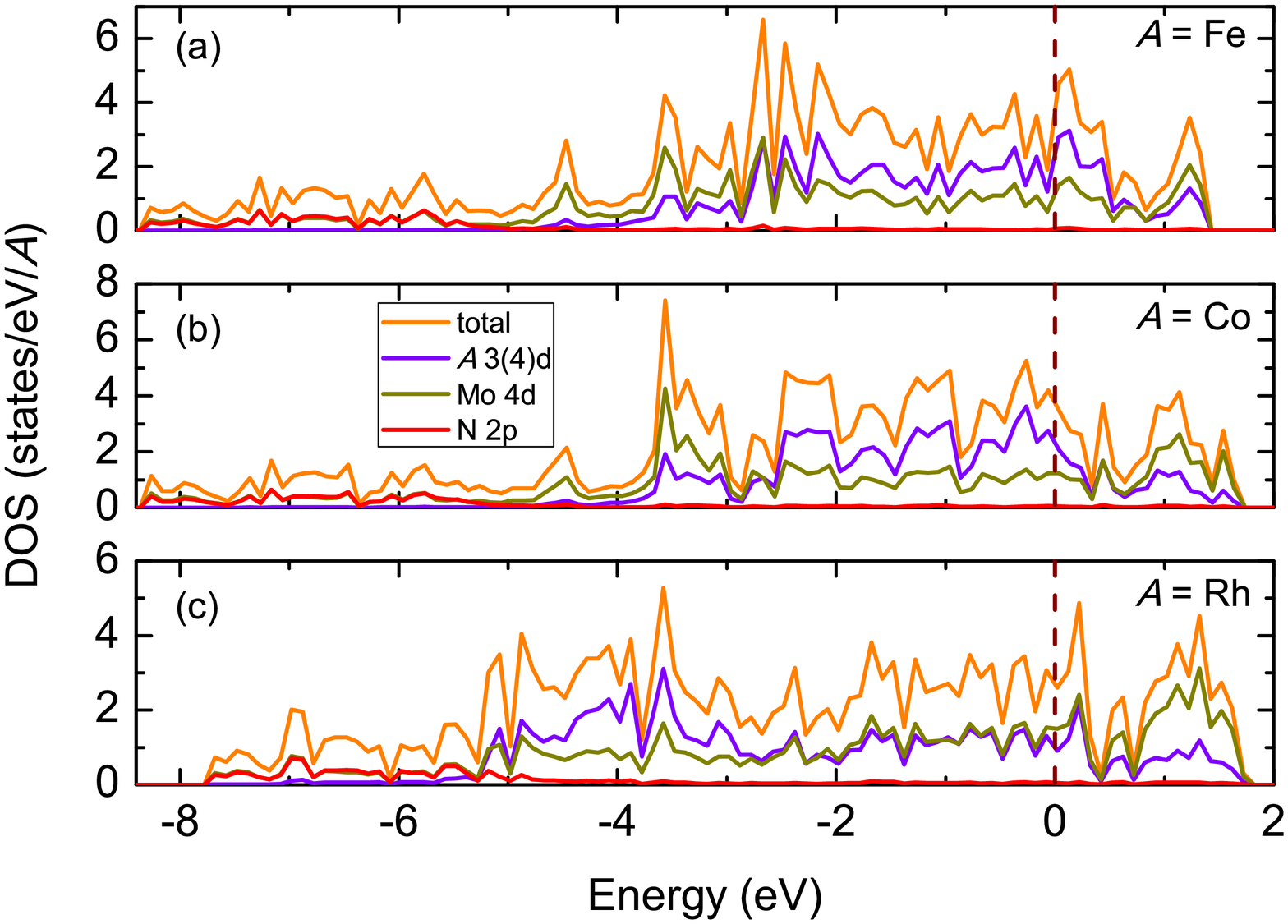}
\caption{(color online) Total DOS and PDOS on $A$ $3(4)d$, Mo $4d$,
and N $2p$ orbitals of the NM state for $A_2$Mo$_3$N with (a)
$A$=Fe, (b) $A$=Co, and (c) $A$=Rh. The Fermi energies are set to
zero.}\label{fig2}
\end{figure}

The electronic properties of $A_2$Mo$_3$N in the quenched
paramagnetic state in $A$ $3(4)d$ and Mo $4d$ orbitals are studied.
Such studies provide references for forthcoming magnetism studies.
By analyzing the density of states (DOS) at the Fermi level we can
infer whether the magnetic state is favored. Fig.~\ref{fig2} shows
the total DOS of $A_2$Mo$_3$N, and its projected DOS (PDOS) onto $A$
$3(4)d$, Mo $4d$, and N $2p$ orbitals respectively. Furthermore, it
also shows that the mixing between $A$ $3(4)d$ and Mo $4d$ occurs
mainly around the Fermi energy ranging from $-4$ eV to $2$ eV
indicating the sizable $3(4)d$-$4d$ hybridization between $A$
$3(4)d$ and Mo $4d$ orbitals. Particularly, comparable contributions
to the conducting carriers are witnessed from the Rh $4d$ and Mo
$4d$ orbitals in Rh$_2$Mo$_3$N, shown in Fig.~\ref{fig2}(c). This is
attributed to the much more extended $4d$ orbitals of Rh compared to
the $3d$ orbitals of Fe and Co, which leads to a stronger
hybridization, expanded bandwith, and highly overlapping between Rh
$4d$ and Mo $4d$ orbital states. Additionally, it is important to
point out that the DOS coming from $A$ $3(4)d$ orbitals is
nonvanishing at the Fermi level. The values of DOS at the Fermi
level are [$N^{Fe}_{Fe}(E_{f}) = 2.4$ and $N^{Fe}_{Mo}(E_{f}) =
0.73$], [$N^{Co}_{Co}(E_{f}) = 2.3$ and $N^{Co}_{Mo}(E_{f}) = 0.8$],
and [$N^{Rh}_{Rh}(E_{f}) = 1.5$ and $N^{Rh}_{Mo}(E_{f}) = 0.67$]
states per eV per $A$(Mo) atom for Fe$_2$Mo$_3$N, Co$_2$Mo$_3$N, and
Rh$_2$Mo$_3$N, respectively. According to the Stoner
criterion~\cite{DJSingh2008,WLi2012}, magnetism may occur only when
the DOS satisfies $N(E_f)I>1$ , where $I$ is the Stoner parameter,
which takes values of $0.7$ eV-$0.9$ eV for ions near the middle of
the transition metal series (note that the effective $I$ can be
reduced by hybridization). It shows clearly that although the NM
state is favorable on the Mo atoms, it is unstable against the
magnetic states on the $A$ atoms. $A$ atoms are responsible for
magnetisms in this family.

To explore the magnetic behaviors of $A$ atoms in $A_2$Mo$_3$N, we
consider a local exchange model based on the nearest neighbor
Heisenberg and DM interactions:
\begin{eqnarray}
H  = \sum_{\langle i,j\rangle}J_{ij}\bS_{i} \cdot \bS_{j}
+\sum_{\langle i,j\rangle}\bD_{ij}\cdot(\bS_{i}\times
\bS_{j}),\label{eq:one}
\end{eqnarray}
where $\bS_i$ is the operator of $A$ spin at site $i$, $\langle
i,j\rangle$ denotes the summation over the nearest neighboring sites
between $A$ atoms. Parameters $J_{ij}$ and $\bD_{ij}$ are the
nearest neighbor Heisenberg and DM interactions, respectively, which
can be evaluated by using the four-state energy-mapping
analysis~\cite{HJXiang}. The results are listed in
Table~\ref{table2}. Here lists only the coupling constants on the
three bonds shown in Fig.~\ref{fig1}(a). Interactions on other bonds
are the same except for a rotation. The vector $\mathbf{D}_{ij}$ of
the DM interaction on each bond are almost perpendicular to the bond
as expected. That is because the lack of inversion symmetry
generates a local electric field $\mathbf{e}$ on each bond. An
electron hopping between two ends of this bond mediates the exchange
interaction of the neighboring spin, and feels an effective magnetic
field $\mathbf{b}=\mathbf{v}\times\mathbf{e}$, where velocity
$\mathbf{v}$ is along the bond. Spin of this electron thus proceed
about $\mathbf{b}$, resulting in a DM interaction of neighboring
spins with the DM vector parallel $\mathbf{b}$. Therefore
$\mathbf{D}_{ij}$ should be perpendicular to the bond. The small
deviation from perpendicular is associated with the non-uniformity
of $\mathbf{e}$. The consistency between this physical picture and
the calculation results justifies our evaluation of the DM
interactions. From these data we notice that the Heisenberg
interactions $J_{ij}$ are almost the same on the three bonds in each
compound. They reach the maximal value in Rh$_2$Mo$_3$N, which
attributes to the nature of strong hybridization between Rh $4d$ and
Mo $4d$ orbitals [see the charge density distribution shown in
Fig.~\ref{fig1}(b)]. The ratio between the DM interaction and
Heisenberg exchange has the largest value
$\gamma=|\frac{\bD_{ij}}{J_{ij}}| \approx$ 0.084 in Co$_2$Mo$_3$N,
while that of Fe$_2$Mo$_3$N and Rh$_2$Mo$_3$N are 0.069 and 0.073,
respectively. These values are much larger than conventional ratios
of $\gamma<0.05$~\cite{moriya_anisotropic_1960}. Such strong DM
interaction between $A$-$A$ atoms in $A_2$Mo$_3$N mainly comes from
strong spin-orbital coupling in Mo $4d$ orbitals, which mediate the
spin-spin interaction between $A$ atoms as indicated from the charge
density distribution shown in Fig.~\ref{fig1}(b). Thus, we conclude
that the Co$_2$Mo$_3$N with strong DM interaction is a promising
candidate for realizing the exotic skyrmion in pure $A_2$Mo$_3$N
series.

\begin{table}
\caption{The values of magnetic exchange coupling and DM
interactions on the three bonds shown in Fig.~\ref{fig1}(a) in the
filled $\beta$-manganese structured $A_2$Mo$_3$N ($A$=Co, Fe, and
Rh). The unit is meV/S$^2$, where S is the spin of the $A$ atom.}
\begin{tabular}{cccc}
\hline
\hline
~~~~~~  & ~~~~~~~$A$=Fe~~~~~~~ &~~~~~~~$A$=Co~~~~~~~ & ~~~~~~~$A$=Rh~~~~~~   \\
\hline
  $J_1$  &-27.17  &-41.46  &-62.71   \\
  $J_2$  &-27.23  &-41.46  &-62.85   \\
  $J_3$  &-27.25  &-41.46  &-62.84   \\
\hline
  $\bD_{1}$&( 0.36,-0.37, 1.80) &( 1.19,-1.16, 3.06) &( 2.39,-2.50, 1.71) \\
  $\bD_{2}$&(-0.53, 0.61,-0.07) &(-1.05, 1.12, 2.69) &(-1.51, 1.26, 4.35) \\
  $\bD_{3}$&( 0.61, 0.48, 0.08) &( 1.11, 1.05,-2.69) &( 2.02, 1.41,-4.31) \\
\hline
\hline
\end{tabular}\label{table2}
\end{table}

As the skyrmion radius is controlled by the ratio between the DM
interaction and Heisenberg exchange, it is of great interesting,
from both physics and applications' perspectives, to find materials
with large $\gamma=|\frac{\bD_{ij}}{J_{ij}}|$. To this end, material
optimizing is required.
Here, we suggest that the substituted compound
Co$_{2-x-y}$Rh$_x$Fe$_y$Mo$_3$N has a larger $\gamma$ value than
pure Co$_2$Mo$_3$N by fine tuning the stoichiometry. According to
the aforementioned calculations and discussions, the strength of DM
can be raised through an enhancement of the hybridization between A
$3(4)d$ and Mo $4d$ orbitals. Therefore partially substituting Co by
Rh will help as the $4d$ orbital of Rh is more extended than the
$3d$ orbital of Co, which makes the overlapping to Mo $4d$ orbital
more sufficient. However, such substitution simultaneously
introduces large atomic orbital potential difference between large
radii Rh and small radii Co atoms, which generates large scattering
on itinerant electrons and weakens the exchange interactions. To
overcome those potential barriers, an isovalent dopant Fe needs to
be introduced. Based on this analysis we further calculate the
magnetic exchange interactions for CoRh$_{0.75}$Fe$_{0.25}$Mo$_3$N
and obtain the ratio $\gamma=|\frac{\bD_{ij}}{J_{ij}}|$ up to
$0.11$. Therefore, we suggest the substituted compound
Co$_{2-x-y}$Rh$_x$Fe$_y$Mo$_3$N is the most promising material
realizing the exotic skyrmion state. An experimental hint has
already been observed in \cite{PDBattle}, where zero remenance and a
kink of magnetic susceptibility both indicate the presence of spin
helices, the forerunner of skyrmions, at small magnetic fields and
low temperatures. A complete exploration of the phase diagram and
stoichiometry is required. On the other hand, the Curie temperature
can be elevated by introducing other dopants such as
Pt\cite{TJPrior2}.

In conclusion, we have constructed a framework of searching new
helimagnet materials harboring exotic spin textures of skyrmions.
The effective Hamiltonian is derived based on symmetry analysis. A
new family $A_2$Mo$_3$N ($A$=Fe, Co, and Rh) has been proposed, and
the first-principle calculations are performed. We hope that this
family is just a corner of a huge iceberg.

We thank H. F. Du, C. L. Chien, M. H. Jiang, X. M. Xie, Z. Liu, and
Y. Li for helpful discussions. WL was supported by the Strategic
Priority Research Program (B) of the Chinese Academy of Sciences
(Grant No. XDB04040300), the National Natural Science Foundation of
China (Grant No. 11227902 and 11404359), and Shanghai Yang-Fan
Program (Grant No. 14YF1407100). JZ was supported by the U.S.
Department of Energy under Award DEFG02-08ER46544, the National
Science Foundation under Grant No. ECCS-1408168, and the Theoretical
Interdisciplinary Physics and Astrophysics Center.

\end{document}